# Generation of five phase-locked harmonics by implementing a divide-by-three optical frequency divider

**Nurul Sheeda Suhaimi[1], Chiaki Ohae[1], Trivikramarao Gavara[1], Ken'ichi Nakagawa[2], Feng Lei Hong[3,4], and Masayuki Katsuragawa[1,4,*]**

[1] *Department of Engineering Science, University of Electro-Communications, 1-5-1 Chofugaoka, Chofu, Tokyo 182-8585, Japan*
[2] *Institute for Laser Science, University of Electro-Communications, 1-5-1 Chofugaoka, Chofu, Tokyo 182-8585, Japan*
[3] *Department of Physics, Yokohama National University, 79-5 Tokiwadai Hodogaya-ku, Yokohama, 240-8501, Japan*
[4] *ERATO, JST, MINOSHIMA Intelligent Optical Synthesizer Project, Honmachi 4-1-8, Kawaguchi, Saitama 332-0012, Japan*
*Corresponding author: katsuragawa@uec.ac.jp

**Abstract:** We report the generation of five phase-locked harmonics, $f_1$: 2403 nm, $f_2$: 1201 nm, $f_3$: 801 nm, $f_4$: 600 nm, and $f_5$: 480 nm with an exact frequency ratio of 1 : 2 : 3 : 4 : 5 by implementing a divide-by-three optical-frequency divider in the high harmonic generation process. All five harmonics are generated coaxially with high phase coherence in time and space, which are applicable for various practical uses.

Frequency division in the optical frequency region has been studied with the aim of coherently linking independent lasers [1-6] as part of a historical trend toward the establishment of an optical frequency standard [7, 8]. Such lasers are attractive not only for establishing a frequency standard but also as a coherent light source itself for various practical uses. Individually, they can be used as single frequency tunable lasers with highly precise frequencies in the frequency domain. As a whole, they can be an excellent candidate for ultrabroad bandwidth coherent light source which in the time domain can be synthesized into ultrashort pulses (< ~fs) with an ultrahigh repetition rate (> ~THz) [1, 9-12]. Furthermore, if the absolute phases of such coherent light source are controlled, we can also synthesize arbitrary optical waveforms (AOWs) [1, 9, 13].

In this *Letter* we demonstrate the generation of five phase-locked harmonics with an exact frequency ratio of 1: 2 : 3 : 4 : 5 by implementing a divide-by-three optical frequency divider [3, 5] in the generation process. We show that the five generated harmonics have relative and absolute phase coherences in time and space that make them suitable for various practical applications including AOW synthesis.

Before proceeding, we briefly review the recent progresses related to the broad phase-locked harmonics generation. Chiodo et al demonstrated a phase-locking of two independent lasers at 1544 nm (Er doped fiber laser) and 1029 nm (Yb doped fiber laser) by matching the third harmonic of the 1544-nm radiation with the second harmonic of the 1029-nm radiation, which resulted in the phase-locked multi frequency radiations ($2f_c$:1544 nm, $3f_c$: 1029 nm, $4f_c$: 772 nm and $6f_c$: 515 nm) [6]. Kung et al demonstrated generation of an automatically phase-locked five harmonics in the nanosecond-pulsed Raman process and based on such harmonics they also showed an amplitude-waveform synthesis in the nanosecond-pulsed regime [13].

We begin our discussion by presenting the concept behind our divide-by-three optical-frequency divider as shown in Fig.1. First, we employ two laser sources as master and divider lasers, whose oscillation frequencies are $f_3$ and $f_2$, respectively. We then produce the difference frequency radiation, $f_1 = f_3 - f_2$, in a difference frequency generation (DFG) process. Furthermore, the $f_1$ radiation is doubled in a second harmonic generation (SHG) process to produce $f_2'\ (=2f_1)$ radiation. Then we detect the RF beat signal between the $f_2$ and $f_2'$ radiations, where the oscillation frequencies of the two laser sources are adjusted in advance so that $f_2 \sim f_2'$, that is, $f_2 \sim \frac{2}{3} f_3$, is satisfied. Finally, we feed back the detected beat signal to the $f_2$ divider- or $f_3$ master-laser source to control their oscillation frequencies so that they exactly satisfy $f_2 = f_2' = \frac{2}{3} f_3$. When this locking is stabilized with a phase coherence between $f_2$ and $f_2'$, the divide-by-three optical-frequency division is completed. We thus obtain the three phase-locked harmonics, $f_1$, $f_2$ and $f_3$ at the exact frequency ratio of 1 : 2 : 3.

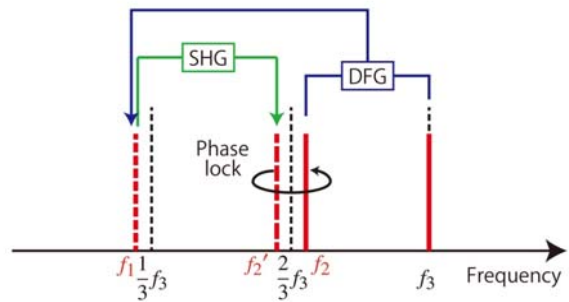

Fig. 1. Concept of a divide-by-three optical-frequency divider

Figure 2 shows the entire system for generating the phase-locked high harmonics, and the principal system elements are the divide-by-three optical-frequency divider and the high harmonic generator. First, we perform divide-by-three optical-frequency division to obtain three phase-locked harmonics, $f_1$, $f_2$ and $f_3$. For this purpose, we employ external cavity diode lasers (ECDLs) oscillating at 801 and 1201 nm for the master ($f_3$) and divider ($f_2$) lasers, respectively. The typical output power of each ECDL is ~ 10 mW. The oscillation frequency of the master 801-nm ECDL is stabilized to a reference cavity (finesse: ~ 2,000) with the Pound-Drever-Hall (PDH) method [14] and is followed by two tapered amplifiers (TA) that can increase the power to 1 W. We couple these 801- and 1201-nm radiations in space with a dichroic mirror and introduce them into a periodically poled lithium niobate waveguide, WG-PPLN1 (length: 48 mm, quasi-phasematching (QPM) pitch: 20.22 μm) to generate the difference frequency, $f_3 - f_2 = f_1$: 2403 nm. This DFG process, which also acts as optical parametric amplifier (OPA), increases the power of 1201-nm radiation simultaneously. We then segregate a small amount of the 2403- and 1201-nm radiations from the main beam line and send them to the divide-by-three optical-frequency divider subsystem. The main beam line is used to generate high harmonics through nonlinear optical mixing processes.

At the divide-by-three optical-frequency divider, we first separate the 2403- and 1201-nm radiations into an individual line. Then we give a frequency shift, -95 MHz, to the 1201-nm radiation using an acoustic optical modulator (AOM). We then recombine the 2403-nm and the frequency-shifted 1201-nm radiations in space and introduce them into another periodically poled lithium niobate waveguide, WG-PPLN2 (length: 48 mm, QPM pitch: 31.3 μm) to generate the second harmonic of $f_1$: 2403-nm radiation, i.e., $f_2'$ radiation. We then heterodyne the generated second harmonic, $f_2'$, and the frequency-shifted 1201-nm, $f_2$ – 95 MHz, radiations in a high-speed photo-diode and detect their RF beat signal, i.e., 95 MHz + ($f_2' - f_2$). We further mix this RF beat signal with a 95-MHz RF standard using a phase detector and generate an error signal. The phase detector employed here has a broad response bandwidth of 200 MHz, which consists of an ultrafast dual comparator (Analog Device AD96687) and a phase and frequency discriminator (Analog Device AD9901). Finally we feed back the obtained error signal to the current of the divider laser, the 1201-nm ECDL; one is through the current controller with a slow feedback of < 100 kHz, the other to the ECDL directly with a fast feedback of < 1.5 MHz. In this way, we achieve divide-by-three optical-frequency division, which results in the generation of three phase-locked harmonics, $f_1$: 2403 nm, $f_2$: 1201 nm and $f_3$: 801 nm at the exact frequency ratio of 1 : 2 : 3.

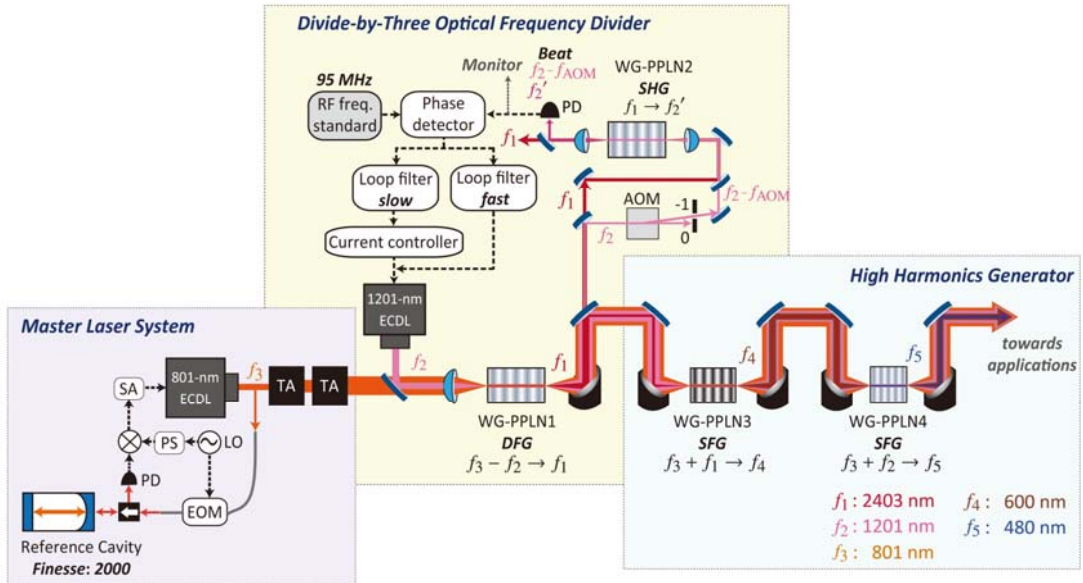

Fig. 2. System layout for generating phase-locked high harmonics. ECDL: external cavity diode laser, TA: tapered amplifier, EOM: electro-optic modulator, LO: local oscillator, PS: phase shifter, SA: servo amplifier, PD: photo detector, WG-PPLN: periodically poled lithium niobate waveguide, AOM: acousto-optic modulator, DFG: difference frequency generation, SFG: sum frequency generation, SHG: second harmonic generation.

Based on these three phase-locked harmonics, we generate further high harmonics through nonlinear optical mixing processes to widen the whole harmonic bandwidth. To achieve this aim, we introduce the three phase-locked harmonics into two periodically poled lithium niobate waveguides, WG-PPLN3 (length: 34 mm, QPM pitch: 11.65 μm) and WG-PPLN4 (length: 23 mm, QPM pitch: 11.63 μm) sequentially to generate $f_4$: 600-nm and $f_5$: 480-nm harmonics by mixing $f_1$: 2403-nm and $f_3$: 801-nm radiations, and $f_2$: 1201-nm and $f_3$: 801-nm radiations, respectively, which can result in the generation of the five phase-locked harmonics with an exact frequency ratio of 1: 2 : 3 : 4 : 5 as a whole. We note that all the harmonics are coupled coaxially from one waveguide to another by using parabolic mirrors (focal length: 15 mm, Ag coated) to avoid chromatic aberration. This coaxial layout of the entire system is practically significant in terms of application, since we can robustly maintain the phase-locked nature among all the harmonics against disturbances. We also note that all of the harmonics, $f_1$ – $f_5$, always maintain the same linear polarization, since we employ the QPM technology for the entire system as shown here.

In the actual operation of the system, the output power of the master ($f_3$: 801 nm) and divider ($f_2$: 1201 nm) lasers were set at 10 and 12 mW, respectively. Then, the output power of the master laser was amplified to 1 W, approximately. At the WG-PPLN1 input, the coupling efficiency of 801-nm and 1201-nm radiations were 92% and 84%, respectively. The DFG conversion efficiency was found to be 421 %·W$^{-1}$. As a result, at the exit of WG-PPLN1, we obtained output powers of 810, 80, and 39 mW for 801-, 1201-, and 2403-nm radiations, respectively. From these output powers, we segregated 2 mW of 1201-nm radiation and 20 mW of 2403-nm radiation for the phase locking. The remaining power was used for the main beam line for generating the phase-locked high harmonics.

First, we evaluated the phase-locking precision and stability of the optical frequency divider. The RF beat signal, $f_2 - f_2'$, was monitored after the WG-PPLN2. Here, the SHG conversion efficiency in the WG-PPLN2 to produce $f_2'$ radiation was around 122%·W$^{-1}$, where the coupling efficiency of the 2403-nm radiation at the WG-PPLN2 input was 43%. Figure 3 shows the frequency spectrum of the beat signal without phase locking (gray), and with phase locking (black). As clearly seen here, when the locking loop was operated, a very sharp peak appeared at exactly 95 MHz (expected beat frequency value) with a locking bandwidth of 1.2 MHz. This bandwidth implies that 84% of the broad beat spectrum in the unlocked case was pulled into the carrier beat frequency at 95 MHz. Thus the carrier beat frequency power became 25 dB greater than the residual noise sideband peaks. These results show that the divide-by-three optical frequency division i.e., the locking of $f_1$, $f_2$, and $f_3$ at the exact integer ratio of 1 : 2 : 3, was achieved.

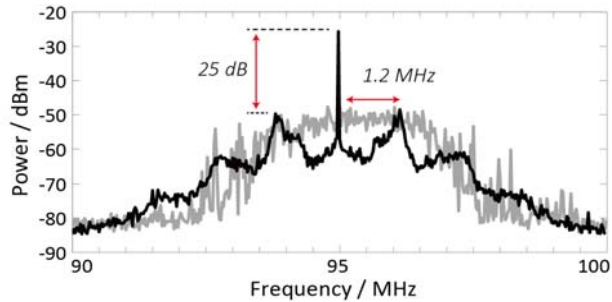

Fig. 3. Frequency spectrum of the beat signal without phase locking (gray) and after phase locking (black).

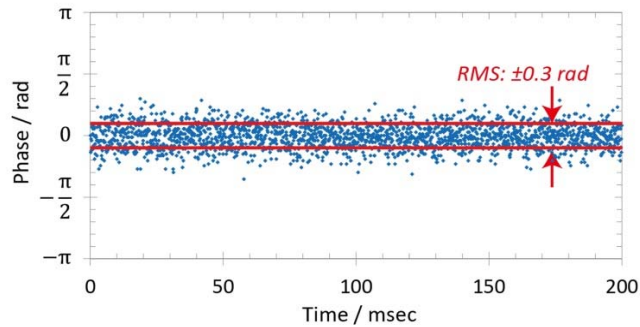

Fig. 4. Phase fluctuations of the beat signal in *a*, short-time and *b*, long-time duration.

To evaluate the precision of this phase-locking condition, we measured the phase fluctuations of the beat signal by using the error signal that was proportional to the phase deviations of the beat signal from the RF frequency standard. Figure 4 shows the phase fluctuations which were measured over a time duration of 200 ms. The mean-square (MS) value of the phase fluctuations was estimated to be 0.096 rad$^2$ at this observation time of 200 msec, which are shown as root-mean-square (RMS) value in Fig. 4. This MS value shows that the power concentration at the exact harmonic frequency was 91% [15], that was also consistent to the observed beat-note spectrum of Fig. 3. This phase locking was maintained stably for at least 30 minutes.

Using these three phase-locked harmonics as a basis, we also generated their high order harmonics, $f_4$ (600 nm ) and $f_5$ (480 nm), by taking sum-frequencies such as $f_1 + f_3$ and $f_2 + f_3$ in the WG-PPLN3 and WG-PPLN4, respectively. At the WG-PPLN3 input, the coupling efficiency of $f_1$: 2403-nm and $f_3$: 801-nm were 66% and 81%, respectively. The conversion efficiency was found to be 149 %·W$^{-1}$. Whereas, at the WG-PPLN4 input, the coupling efficiency of $f_2$: 1201 nm and $f_3$: 801 nm were 54% and 57%, respectively, with the conversion efficiency of 235 %·W$^{-1}$. Figures 5a, and b show photographs of the five harmonics obtained in this way (except $f_1$: 2403 nm), which were taken with a CCD camera (a) before and (b) after we dispersed the beam with a prism onto a white screen. As seen here, all five harmonics were generated coaxially with a Gaussian-like round beam profile. The respective output powers were 0.8, 33.5, 341.7, 5.5 and 29.5 mW for $f_1$: 2403, $f_2$: 1201, $f_3$: 801, $f_4$: 600 and $f_5$: 480 nm at the exit of WG-PPLN4. We also measured the beam profiles for the generated high order harmonic components. $f_4$: 600 nm and $f_5$: 480 nm harmonics exhibited a good beam quality, namely an M$^2$-value of better than 1.2. The bandwidth of the phase-locked harmonics was expanded from 374.2 THz ($f_1 - f_3$: $2403\ nm - 801\ nm$) to 623.7 THz ($f_1 - f_5$: $2403\ nm - 480\ nm$).

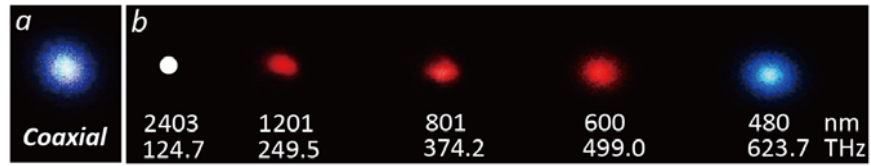

Fig. 5. Photos of the five generated phase-locked harmonics, *a*, before, *b*, after dispersal of the harmonic beam with a prism. The 2403-nm radiation is shown by a solid white circle.

Finally, to determine the potential of the five produced harmonics for various practical applications, we performed an interference measurement, namely, we observed the interferences among the five harmonics via nonlinear optical processes. Figure 6 shows a typical result. The sum frequency component (named $f_7^\beta$) was generated in two ways in a thin BBO crystal (Type I phasematching, thickness: 10 μm), namely SFGs via $f_2 + f_5$ and $f_3 + f_4$, and they were superimposed onto a photo detector (see the inset in Fig. 6). Furthermore, the spectral phases of these five harmonics were manipulated by inserting a pair of Calcium Fluoride plates (thickness: 5 mm) in the optical path after WG-PPLN4 and tilting them symmetrically with exactly the same angle of α (this phase manipulation is described in detail in Refs. 12, 16). The sinusoidal-like interference pattern was clearly observed, and was well fitted with a theoretically predicted curve, which shows that the phase relationship among the high harmonics was well maintained in time and space including its absolute phase. Furthermore, these harmonic frequencies were tunable by tuning the frequency of the 801 nm master laser. Even though the master laser frequency was tuned around 1 GHz, the phase-locked condition was maintained. The produced five phase-locked harmonics with an exact frequency ratio of 1 : 2 : 3 : 4 : 5, are applicable to AOW synthesis in the time domain and also to single frequency tunable lasers in the frequency domain.

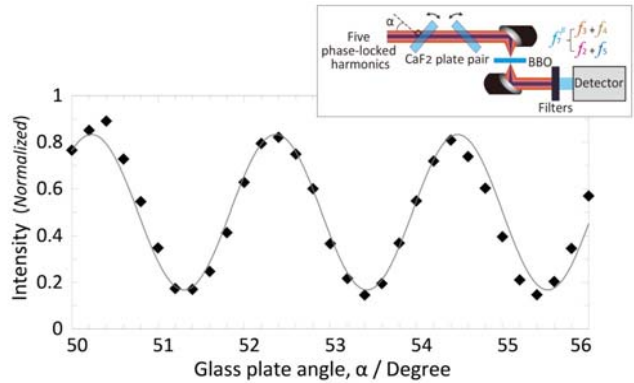

Fig. 6. Interference signal aroused with two superimposed sum-frequency components, $f_2 + f_5$ and $f_3 + f_4$, as a function of the angle α of a glass plate pair inserted in the optical path.

In summary, we have demonstrated the generation of five phase-locked harmonics, $f_1$: 2400 nm, $f_2$: 1201 nm, $f_3$: 801 nm, $f_4$: 600 nm, and $f_5$: 480 nm with an exact frequency ratio of 1 : 2 : 3 : 4 : 5, by implementing a divide-by-three optical frequency divider in it. The output powers we finally obtained were $f_1$: 0.8, $f_2$: 33.5, $f_3$: 341.7, $f_4$: 5.5 and $f_5$: 29.5 mW, respectively. All these five phase-locked harmonics were generated coaxially and exhibited high phase coherence with each other in time and space, which was sufficient for them to be applicable to AOW synthesis in the time domain, including ultrashort pulse waveform (1.8-fs pulse duration [12, 16]) and also to ultrahigh-precision single-frequency tunable lasers in the frequency domain.

We note that the frequency spacing of the produced harmonics exactly matches the vibrational Raman transition ($v = 1 \leftarrow 0$: 125 THz) in gaseous para-hydrogen. Therefore, they can be used to generate high order stimulated Raman scattering [17-19] which can extend the harmonics wavelength range to the extreme region (mid infrared to vacuum ultraviolet). We also note that the technology studied here can be extended to a frequency comb of ultrashort pulses to efficiently expand the entire bandwidth of the frequency comb.

**Funding.** Grant-in-Aid for Scientific Research (A), No. 24244065.

**Acknowledgment.** We thank N. Watanabe and A. Suzuki for their supports in setting up the system. We also acknowledge Y. Nishida for providing knowledge regarding the periodically poled lithium niobate waveguide.